\documentclass[10pt,aip,jap,superscriptaddress,twocolumn,floatfix]{revtex4-2}

\usepackage{graphicx}
\usepackage{xcolor}
\usepackage{url} 
\usepackage{hyperref} 
\usepackage{dcolumn}
\usepackage{bm}
\usepackage{multirow}
\usepackage[symbol]{footmisc}
\usepackage{amsmath}
\usepackage{amssymb}
\usepackage{amsfonts}
\usepackage{subfigure}
\usepackage{url}
\usepackage{soul}

\begin{document}

\title{Magnetic field-induced chiral soliton lattice in the bulk magnetoelectric helimagnet Cu$_2$OSeO$_3$}

\author{Victor Ukleev}
\affiliation{Laboratory for Neutron Scattering and Imaging (LNS), Paul Scherrer Institute (PSI), CH-5232 Villigen, Switzerland}
\affiliation{Helmholtz-Zentrum Berlin f\"ur Materialien und Energie, D-14109 Berlin, Germany}
\email{victor.ukleev@helmholtz-berlin.de}
\author{Arnaud Magrez}
\affiliation{Crystal Growth Facility, Institute of Physics, \'Ecole Polytechnique F\'ed\'erale de Lausanne, CH-1015 Lausanne, Switzerland}
\author{Jonathan S. White}
\affiliation{Laboratory for Neutron Scattering and Imaging (LNS), Paul Scherrer Institute (PSI), CH-5232 Villigen, Switzerland}

\date{\today}

\begin{abstract}
Chiral soliton lattices (CSLs) are anharmonic magnetic structures typically found in uniaxial chiral magnets. In this study, we report the observation of CSL in bulk Cu$_2$OSeO$_3$, a chiral insulator known for its magnetoelectric properties. Using small-angle neutron scattering (SANS) experiments, we demonstrate the formation of CSLs in Cu$_2$OSeO$_3$ at low temperatures, driven by the competition between cubic anisotropy and magnetic field. Our observations of higher harmonics in the SANS signal clearly indicate the anharmonic nature of the spiral. This finding underscores the complex interplay between magnetic interactions in Cu$_2$OSeO$_3$, offering insights for potential applications of CSLs in electric-field controlled spintronic devices.
\end{abstract}

\maketitle

\section{Introduction}
\begin{figure*}
\includegraphics[width=17cm]{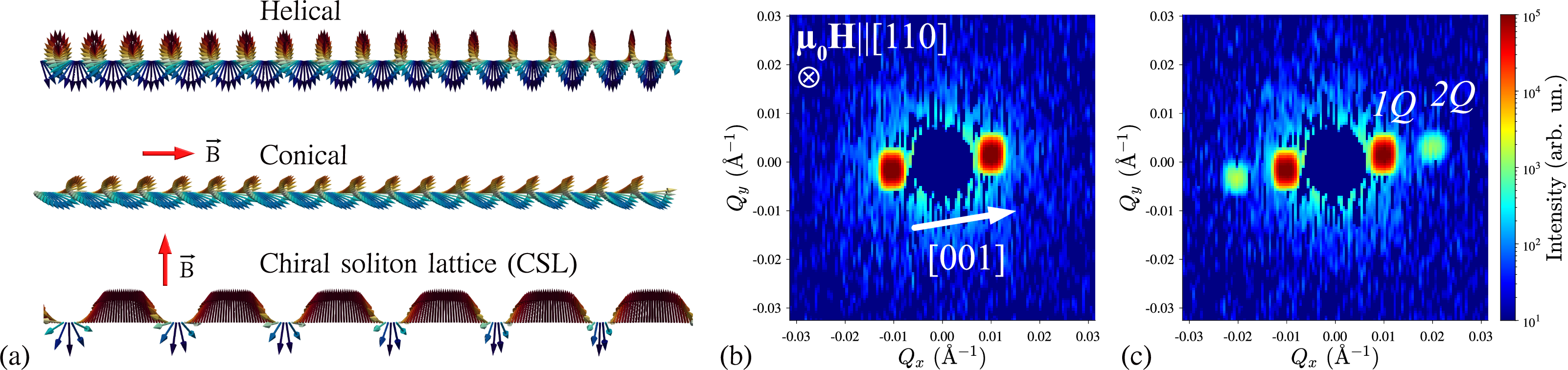}
\caption{(a) Schematic illustrations of spin patterns in the helical state, conical, and chiral soliton lattice (CSL) states. (b,c) Magnetic small-angle neutron scattering (SANS) patterns obtained at $T=2$\,K and (b) zero field and (c) $\mu_0 H=22$\,mT. First-order SANS peaks $1Q$ are accompanied by the higher-order satellites $2Q$ due to the field-induced deformation of the spiral.}
\label{fig1}
\end{figure*}

The interplay of exchange and Dzyaloshinskii-Moriya interactions (DMI) and the magnetocrystalline anisotropy yields a diverse magnetic phase diagram in chiral cubic crystals, encompassing paramagnetic, helimagnetic, forced ferromagnetic, conical, and skyrmion phases \cite{tokura2020magnetic}. Application of a magnetic field parallel to the propagation vector of a magnetic helix results in the canting of the magnetic moments towards the field direction, leading to the transformation of the helical texture into a conical state (Fig. \ref{fig1}a). Further increments of the magnetic field result in the transition from the conical to the forced ferromagnetic state. If significant cubic anisotropy is present, a moderate magnetic field applied perpendicular to the helical propagation axis can deform the proper screw magnetic modulation into an elliptic spiral \cite{dzyaloshinskii1965theory,izyumov1984modulated} also known as chiral soliton lattice (CSL) \cite{togawa2012chiral} (Fig. \ref{fig1}a). However, the CSL state typically appears in uniaxial chiral magnets due to the interplay between easy-axis anisotropy and external magnetic fields \cite{togawa2012chiral}, and this distorted state is often neglected for cubic systems, despite its signature has been clearly observed in the prototype $B20$ compound MnSi \cite{grigoriev2006magnetic,kousaka2014chiral}. Another peculiar, multi-$k$ magnetic solitonic texture that arises in chiral helimagnets under applied magnetic fields is a skyrmion crystal, a hexagonally ordered array of topologically protected chiral magnetic vortices \cite{muhlbauer2009skyrmion,yu2010real}. The cubic anisotropy affects the shape of skyrmions \cite{ukleev2019element,preissinger2021vital} and the skyrmion lattice orientation with respect to applied field \cite{adams2018response}. Furthermore, the competition between cubic and anisotropic exchange anisotropies enriches the magnetic phase diagram with tilted conical and low-temperature disordered skyrmion phases \cite{chacon2018observation,qian2018new,bannenberg2019multiple,baral2023direct}. 

Magnetic solitons hold promise as information carriers in next-generation non-volatile memory and logic devices \cite{kim2017switching}. Despite extensive theoretical investigations into the current-induced motion of CSLs in itinerant chiral magnets \cite{bostrem2008theory,koumpouras2016spin}, the influence of electric ($E-$) fields has not been explored, although the formation of CSLs in insulating materials has been reported \cite{schefer2002soliton}. The creation and manipulation of CSLs by $E-$fields offer an alternative means of controlling CSLs in insulating helimagnets.

The ground state of bulk Cu$_2$OSeO$_3$~is a proper-screw spiral with the propagation vector parallel to $\langle 100 \rangle$ crystal axes, resulting in a multidomain zero-field state. Upon the application of a magnetic field, the system undergoes a transition from the helical to the conical or skyrmion lattice phase \cite{seki2012observation}. Recent experiments have demonstrated CSL stabilization in $B20$-type magnetoelectric Cu$_2$OSeO$_3$~by applying uniaxial stress or tensile strain \cite{okamura2017emergence,nakajima2018uniaxial}. Furthermore, By exploiting the magnetoelectric nature of Cu$_2$OSeO$_3$, previous studies have demonstrated the $E-$field-induced phase control of chiral spin textures, such as skyrmion lattices, helical, and conical orders in bulk samples and thin plates using moderate $E-$fields of 5-30\,kV/mm \cite{white2012electric,white2014electric,ruff2015magnetoelectric,milde2016heuristic,okamura2017directional,white2018electric,huang2018situ,huang2022electric,han2023hysteretic}.

In the present paper, using small-angle neutron scattering (SANS), we demonstrate magnetic field-induced CSL formation in Cu$_2$OSeO$_3$ at low temperatures due to the cubic anisotropy and its manipulation by magnetic fields.
Although most theoretical treatments of CSL formation focus on uniaxial systems, our observations show that even modest cubic anisotropy is sufficient to stabilize the CSL phase. To our knowledge, no explicit theoretical framework currently addresses this mechanism in cubic systems; however, phenomenological models such as the sine-Gordon description remain applicable. Using the observed critical field for the spiral reorientation ($\sim40$,mT), one can estimate the anisotropy strength required to overcome the Zeeman energy, providing a rough but self-consistent validation of this mechanism in Cu$_2$OSeO$_3$. The obtained magnetic-field dependence of the modulation vector along the [001] direction is in good agreement with the sine-Gordon theoretical model. This makes Cu$_2$OSeO$_3$~a promising material for the non-volatile manipulation of CSLs.

\section{Experimental}

For our experiments, a single crystal of Cu$_2$OSeO$_3$ was grown using a chemical vapor transport method \cite{seki2012observation}. We performed SANS experiments on a $3\times3\times1$\,mm$^3$ sample with the thin axis parallel to the cubic [001] direction.

The crystal was mounted onto a dedicated sample stick \cite{bartkowiak2014note} and loaded into a 7\,T horizontal-field cryomagnet MA7 (Oxford Instruments, UK). We utilized a longitudinal SANS configuration with the incoming neutron's wavevector $k_i||\mathbf{H}||[110]$ (Fig. \ref{fig1}b). 

The SANS measurements were conducted using the SANS-I setup at the Swiss Spallation Neutron Source SINQ, Paul Scherrer Institute (Villigen, Switzerland). We employed an incident neutron beam with the wavelength $\lambda=5$\,\AA~and the bandwidth of 10\%, collimated over a distance of 18\,m before the sample, and the diffracted neutrons were collected by a position-sensitive detector (PSD) located 18\,m after the sample. The SANS patterns were measured by rocking the sample and cryomagnet together through an angular range of $\pm 4^\circ$ that moved the diffraction spots through the Bragg condition at the detector with the step size of $0.5^\circ$. Background measurements were carried out in the field-polarized state at $\mu_0 H = 150$\,mT and subtracted from the other datasets to emphasize the signal due to the magnetic modulations. The data was analyzed using the GRASP software \cite{dewhurst2023graphical}.

\section{Results and Discussion}

\begin{figure*}
\includegraphics[width=17cm]{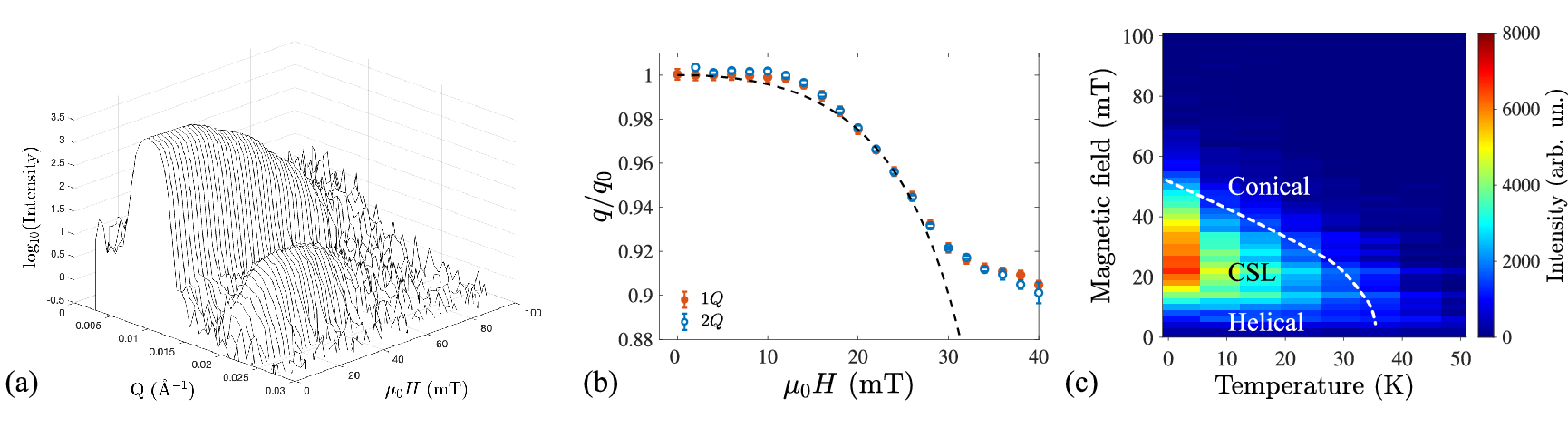}
\caption{(a) Waterfall plot of the integrated intensities of the SANS pattern measured at $T=2$\,K. Appearance of the second-harmonic peak at $Q\approx0.02$\,\AA$^{-1}$ indicates the CSL state. (b) The magnetic-field dependence of the normalized wavevector extracted from the fitted $Q_1$ and $Q_2$ peak positions. The dashed line is the fit according to the sine-Gordon model. (c) Colormap plot of the second harmonic peak intensity corresponding to the order parameter of the CSL state. Dashed line indicates the approximate boundary between the CSL and conical phases, based on the sharp drop in the $2Q$ harmonic intensity.}
\label{fig2}
\end{figure*}

Figure \ref{fig1}b shows the SANS measured at $T = 2$\,K and zero magnetic field after zero-field cooling (ZFC). The diffraction pattern comprises two Bragg peaks from the helical magnetic texture aligned along the [001] axis. In this geometry, diffraction arising from the other two helical domains is out of the scattering plane. Consistent with previous studies, the length of the magnetic wavevector at zero field is $Q_0=0.0098(1)$\,\AA$^{-1}$, corresponding to a real-space periodicity of 64\,nm. The width of the Bragg peaks is limited by the experimental resolution. The zero-field proper-screw helical state has a purely sinusoidal magnetic structure factor that results in the single Fourier harmonic seen in the SANS pattern. Upon increments of the magnetic field, the spiral acquires the net magnetization along the field direction through the elliptic deformation. Thus, the corresponding Fourier transform of the magnetization (SANS pattern) shows higher harmonics \cite{izyumov1984modulated}. 

Experimentally, this transformation from helical to anharmonic CSL upon the field increase is evidenced in SANS, where even a small field of 4\,mT is sufficient to generate the anharmonicity and, consequently, the emergence of the second harmonic (Fig. \ref{fig1}c). This anharmonic spiral is distinct from other modulated magnetic states in Cu$_2$OSeO$_3$, such as the conical or skyrmion lattice phases, by the presence of higher-order harmonics in the SANS pattern and the absence of hexagonal symmetry. Observation of even higher harmonics would also be possible with a bigger sample or longer SANS acquisition times. In order to resolve fine details of the helicoidal structure in the real-space, a high-resolution coherent resonant x-ray scattering \cite{ukleev2018coherent,tabata2020observation} or Lorentz transmission electron microscopy \cite{togawa2012chiral} would be required. In contrast to the conical state, where a single $Q$-vector aligned with the magnetic field dominates, the CSL retains a nontrivial propagation vector orientation and a multi-harmonic structure, even within the coexistence range around 30–40\,mT. The transition from helical to CSL textures upon the field increase progresses up to the point when the magnitude of the field is sufficient for the re-orientation of the spiral wavevector towards the field axis (conical state). At $\mu_0 H\approx40$\,mT, the CSL almost fully transforms into the conical state with the wavevector aligned towards the magnetic field direction, which is not contributing to the SANS signal in the present geometry. The magnitude of this critical field corresponds to the energy barrier of the spiral plane flop provided by the magnetocrystalline anisotropy.

The magnetic-field dependence of the radially integrated SANS intensities at $T=2$\,K is shown in Fig. \ref{fig2}a. The intensity dependence of the second harmonic at $Q\approx0.02$\,\AA$^{-1}$ generally follows predictions of the sine-Gordon model \cite{izyumov1984modulated}, in which the higher-harmonic intensities gradually increase with the applied magnetic field, and with the previous experimental data on the strained Cu$_2$OSeO$_3$ \cite{okamura2017emergence} and FeGe \cite{ukleev2020metastable} nanostructures, and bulk MnSi \cite{grigoriev2006magnetic,kousaka2014chiral}. 

The magnetic-field dependence of the CSL normalized wavevector $q/q_0$, has been derived from the Gaussian fit of the integrated intensities of the SANS patterns (Fig. \ref{fig2}b). The dependence follows the sine-Gordon law (dashed line in Fig. \ref{fig2}b) similarly to uniaxial chiral magnets if the demagnetization effect is taken into account \cite{honda2020topological}.
However, when the field is above $\sim30$\,mT, and the CSL co-exists with the out-of-plane conical state, the $Q-$dependence clearly deviates from the theoretical trend. At this field, the anharmonic spiral with the propagation vector along [100] stabilized by the magnetocrystalline anisotropy becomes less energy favorable than the field-aligned conical state. Therefore, only metastable CSL domains survive the transition field of 30\,mT, but further changes in their shapes becomes impossible and results in the spiral plane flop.

Finally, the intensity of the second harmonic peak $2Q$ is summarized in a temperature vs. magnetic field map as shown in Fig. \ref{fig2}c. The intensity is peaked at the base temperature 2\,K and the field of $22$\,mT, and gradually vanishes at $T\sim40$\,K. Such behavior was also observed in bulk MnSi \cite{grigoriev2006magnetic} and explained by the presence of the cubic anisotropy \cite{maleyev2006cubic}. This observation also aligns well with the temperature behavior of magnetocrystalline \cite{halder2018thermodynamic} and exchange \cite{baral2023direct} anisotropies in Cu$_2$OSeO$_3$.   

\section{Conclusion}

In summary, our investigation delves into the stability of the chiral soliton lattice (CSL) within bulk unstrained Cu$_2$OSeO$_3$. Utilizing small-angle neutron scattering experiments, we unveil the magnetic field-induced emergence of CSLs at low temperatures, attributed to magnetocrystalline anisotropy. The observation of higher-harmonic satellites underscores the significant anharmonicity of the helical spiral. Interestingly, Kelvin-probe force microscopy measurements \cite{milde2016heuristic} indicate the development of a finite electric polarization along the $\langle 100 \rangle$ axis at small magnetic fields applied along $\langle 110 \rangle$. This axis corresponds to the direction of the helical wavevector in the CSL state, suggesting that the observed polarization may originate from a non-collinear magnetoelectric coupling unique to the anharmonic CSL structure as it was also proposed in other magnetoelectric chiral helimagnets \cite{miyake2015magnetic,araki2020metamagnetic}. This highlights the complexity of the interplay between electric polarization and magnetic fields and calls for further theoretical and experimental investigations to elucidate the role of anisotropic magnetoelectric coupling in shaping the CSL dynamics in Cu$_2$OSeO$_3$.

\section*{Acknowledgements}

Authors thank O.I. Utesov for enlightening discussions. We thank M. Bartkowiak for the experimental support. V.U., A. M., and J.S.W. acknowledge funding from the SNSF Sinergia CRSII5-171003 NanoSkyrmionics. J.S.W. also acknowledges funding from the SNSF Project 200021\_188707. SANS measurements at the SINQ (Paul Scherrer Institut) were performed as a part of the proposal 20180048. 

\section*{Author Declarations}
\subsection*{Conflict of Interest}
The authors have no conflicts to disclose.
\subsection*{Author Contributions}
V.U., J.S.W., performed experiments, A.M. synthesized the sample, V.U. analyzed the data, V.U., J.S.W. wrote the manuscript, V.U., and J.S.W. jointly conceived the project. All authors read and edited the manuscript.
\section*{Data availability Statement}
Data is available from Zenodo repository \onlinecite{data2024psi}.  

%

\end{document}